\documentclass{osa-article}

\journal{empty}

\usepackage{wrapfig}
\usepackage{subcaption}
\usepackage{diagbox}
\usepackage{textcomp}
\articletype{Research Article}

\begin{document}

\title{Random Walk for modelling Multi Core Fiber cross-talk and step distribution characterisation} 

\author{Alessandro Ottino, Hui Yuan, Yunnuo Xu, Eric Sillekens, Georgios Zervas}
\address{Optical Networks Group, Dept. Electronic \& Electrical Engineering, UCL, London WC1E 7JE, U.K.}
\email{alessandro.ottino.16@ucl.ac.uk}

\begin{abstract}
A novel random walk based model for inter-core cross-talk (IC-XT) characterization of multi-core fibres capable of accurately representing both time-domain distribution and frequency-domain representation of experimental IC-XT has been proposed. It was demonstrated that this model is a generalization of the most widely used model in literature to which it will converge when the number of samples and measurement time-window tend to infinity. In addition, this model is consistent with statistical analysis such as short term average crosstalk (STAXT), keeping the same convergence properties and it showed to be almost independent to time-window. To validate this model, a new type of characterization of the IC-XT in the dB domain (based on a pseudo random walk) has been proposed and the statistical properties of its step distribution have been evaluated. The performed analysis showed that this characterization is capable of fitting every type of signal source with an accuracy above 99.3\%. It also proved to be very robust to time-window length, temperature and other signal properties such as symbol rate and pseudo-random bit stream (PRBS) length. The obtained results suggest that the model was able to communicate most of the relevant information using a short observation time, making it suitable for IC-XT characterization and core-pair source signal classification. Using machine-learning (ML) techniques for source-signal classification, we empirically demonstrated that this technique carries more information regarding IC-XT than traditional statistical methods.   
\end{abstract}


\section{Introduction}
Space division multiplexing (SDM) technology employing multi-core fiber (MCF) is considered one of the most promising candidates to meet the capacity demands for future optical fiber communication systems \cite{Richardson_2013}. However, MCF-based communication system could be limited/impaired by the inter-core cross-talk (IC-XT) between adjacent cores, which can reduce optical signal-to-noise ratio (OSNR), and therefore system performance and power budget \cite{Puttnam_2016}. For this reason, the analysis of the IC-XT and its statistical properties is crucial for the development of future SDM networks. Different studies have been carried out to characterize IC-XT \cite{Hayashi_2014,Hayashi_2017,Alves_2016,Alves_2018}, among which the model presented in \cite{Hayashi_2014} has been widely used.

\subsection{Time-Domain Distribution} \label{subsec-Time}
\begin{figure}[t]
  \begin{minipage}[b]{0.49\textwidth}
  \centering
        \includegraphics[width=\textwidth]{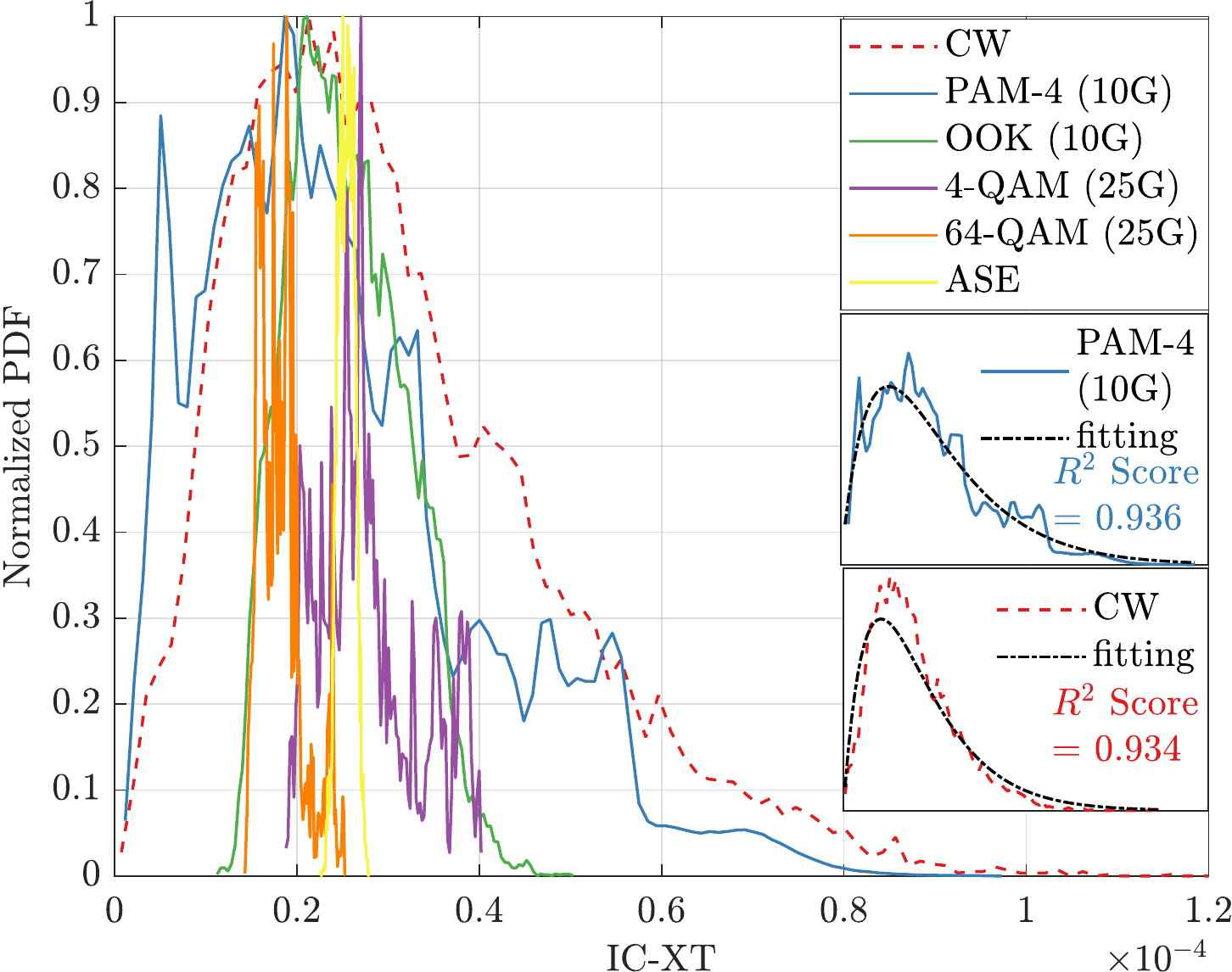}
        \captionof{figure}{Normalized IC-XT PDFs and \\
        $\chi^2$ fitting}
        \label{fig:chi_fit}
    \end{minipage}
\hfill
  \begin{minipage}[b]{0.49\textwidth}
  \centering
    \footnotesize
    \begin{tabular}{c|cc}
        \hline
        Source signal & \multicolumn{1}{c|}{\textbf{Fitt. Acc}} & \textbf{$\sigma$} \\[-0.5ex] \hline 
         CW & \multicolumn{1}{c|}{93.37\%} & 0.00278419 \\[-0.4ex]
        ASE & \multicolumn{1}{c|}{-113.83\%} & 0.00252016 \\[-0.4ex]
        OOK 10G & \multicolumn{1}{c|}{40.38\%} & 0.00282615 \\[-0.4ex]
        PAM4 & \multicolumn{1}{c|}{93.60\%} & 0.00252016 \\[-0.4ex]
        QAM256 & \multicolumn{1}{c|}{-27.59\%} & 0.00183112 \\[-0.4ex] \hline \hline
        Time Window & \multicolumn{2}{c}{Fitness Accuracy} \\[-0.4ex] \cline{2-3} 
         & \multicolumn{1}{c|}{\textbf{CW}} & \textbf{PAM4} \\[-0.5ex] \hline
        12 hours & \multicolumn{1}{c|}{93.37\%} & 93.60\% \\[-0.4ex]
        6 hours & \multicolumn{1}{c|}{87.26\%} & 77.19\% \\[-0.4ex]
        1 hour & \multicolumn{1}{c|}{61.40\%} & -0.22\% \\[-0.4ex] \hline
    \end{tabular}
    \captionof{table}{$\chi^2$ fitting accuracy with different\\ source signals and time windows}
    \label{tab:fit_chi}
  \end{minipage}
 \end{figure}

In \cite{Hayashi_2014}, IC-XT is presented as the sum of four squared Gaussian distributed random variables, making its probability density function (PDF) follow a $\chi^2$ distribution with four degrees of freedom as follows:
\begin{align}
    \label{chi_dist} f_{\chi^2, 4df}(x|\sigma) = \frac{x}{4\sigma^4}e^{\frac{-x}{2\sigma^2}}.
\end{align}This statistical behaviour has been experimentally validated for CW and OOK signals \cite{Hayashi_2014, Luis_2016}. Our independent experimental analysis in \cite{Yuan_2019, yuan2020arxiv} (data in \cite{Dataset:measurements}) showed an agreement with the above equation for PAM4 and CW signal sources, both having a $\chi^2$ fitting accuracy of $>$93\% over 12-hours of results, evaluated through $R^2$ score function \cite{COLINCAMERON1997329} (shown in Fig.\ref{fig:chi_fit}). 

 Twelve hours long measurements are used as a benchmark due to the low divergence in the standard deviation for CW and PAM-4 source signals when the time window increases from 10 to 12 hours, having an average divergence of 0.7\% with ninety percent confidence interval (CI$_{90\%}$) between 0.24\% and 1.3\%.  This suggests that twelve hours observation time contains enough information to perform statistical analysis. However, IC-XT  for other source signals clearly differed from the aforementioned behaviour (Table \ref{tab:fit_chi}). Moreover, when the observation time varies, the shape of the PDF of the data changes drastically (Table \ref{tab:fit_chi}). For example, for PAM4, halving the time leads to a $R^2$ score accuracy reduction of 16 percentage points  and using a time window of one hour, we have a fitness accuracy of -0.22\%. For this reason, the previously mentioned statistical methods cannot be considered reliable when short term real-time data (minutes-hours) are required for dynamic applications such as source signal characterisation and classification, dynamic crosstalk mitigation, optical signal attack detection, etc..

\subsection{Frequency Domain Representation}
\label{subsec:Freq}
In \cite{Hayashi_2014}, the cross-talk between target core n and active core m of a homogeneous MCF was represented through the following equation:
\begin{align}
    A_n(N_{\text{PM}}) &= A_n(0) -j \sum_{l=1}^{N_{\text{PM}}} \chi_{nm}(l)\text{exp}\left[-j \phi_{\text{rnd}(l)}\right]A_m(l-1)\\
    &\approx -j \sum_{l=1}^{N_{\text{PM}}} \chi_{nm} \text{exp}(-j \phi_{\text{rnd},l}) \text{  Where: }\phi_{\text{rnd},l} \backsim U(0, 2\pi)
\end{align}
where $A_n$ is the complex amplitude of the IC-XT of the target core n, $A_m$ is the complex amplitude of the signal in the active core, $\chi_{nm}$ is the coupling coefficient between cores n and m, $\phi_{\text{rnd},l}$ is the random phase shift (0 to 2$\pi$) at the $l^{th}$ phase matching point (PMP) and $N_{\text{PM}}$ is the total number of phase matching points between cores n and m in the MCF.
In the model described in \cite{Hayashi_2014}, $\phi_{\text{rnd},l}$ varies randomly with time, causing the cross-talk to be considered as a random variable sampled from the distribution described in Eq.\ref{chi_dist}. This leads to uncorrelated IC-XT samples. However, statistical analysis of the IC-XT, in terms of auto-correlation and auto-covariance, showed that IC-XT samples which were closer in time are more correlated to each other \cite{Alves_2018}.
In addition, in the frequency domain, sampling from a random variable would lead to a DC bias with constant amplitude for all other frequency components. This largely differs from the frequency domain representation of our experimental data. For both the auto-correlation and frequency domain behaviours, the cross-talk follows a similar trend to the one of a random walk, i.e, a cumulative sum of stochastic random variables. 
\begin{figure}[t]
  \begin{minipage}[b]{0.49\textwidth}
    \centering
    \includegraphics[width=\linewidth]{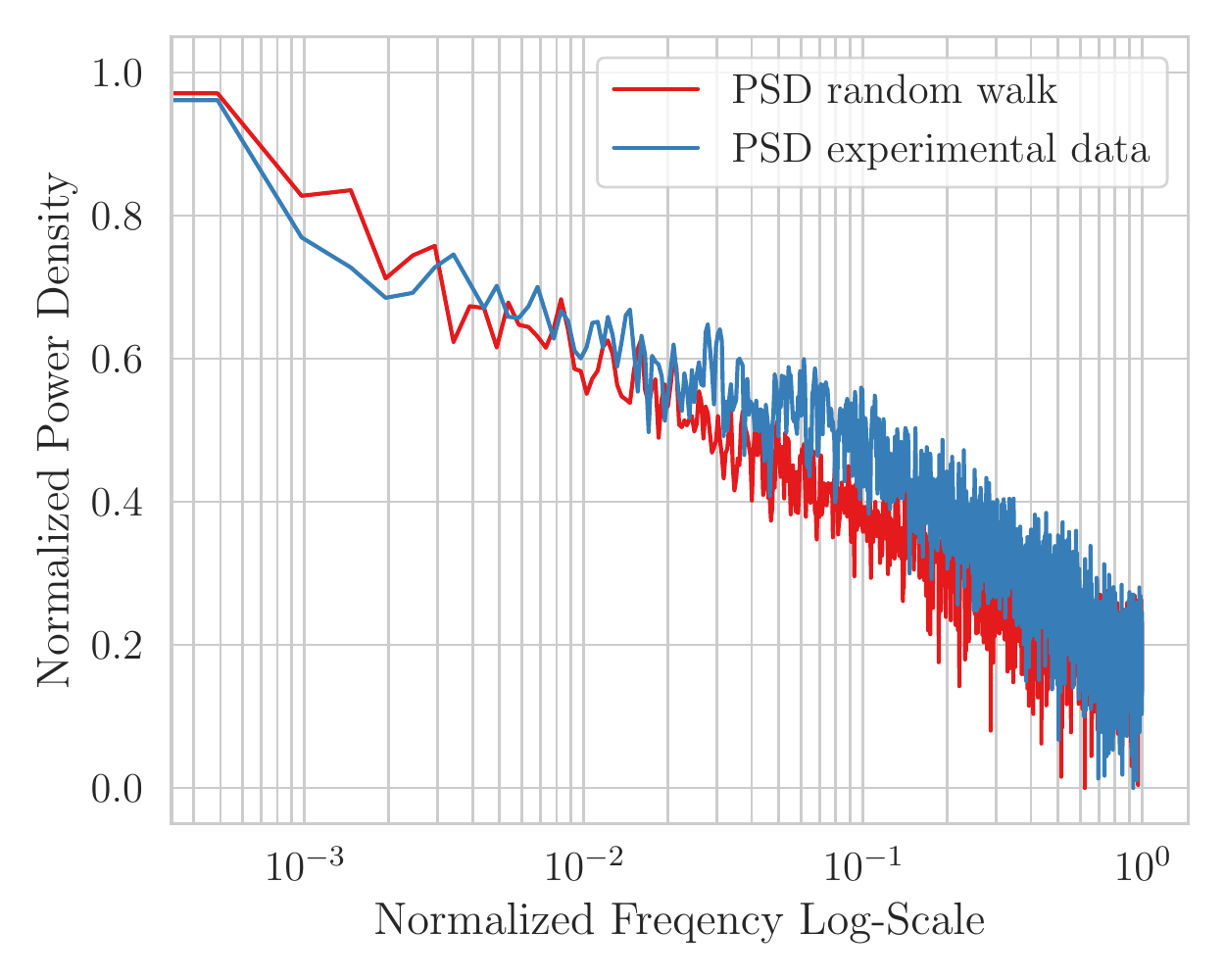}
    \caption{Normalized PSD in log scale of random\\
    walk and experimental IC-XT}
    \label{fig:PSD}
  \end{minipage}
  \hfill
  \begin{minipage}[b]{0.49\textwidth}
    \centering
    \includegraphics[width=\linewidth]{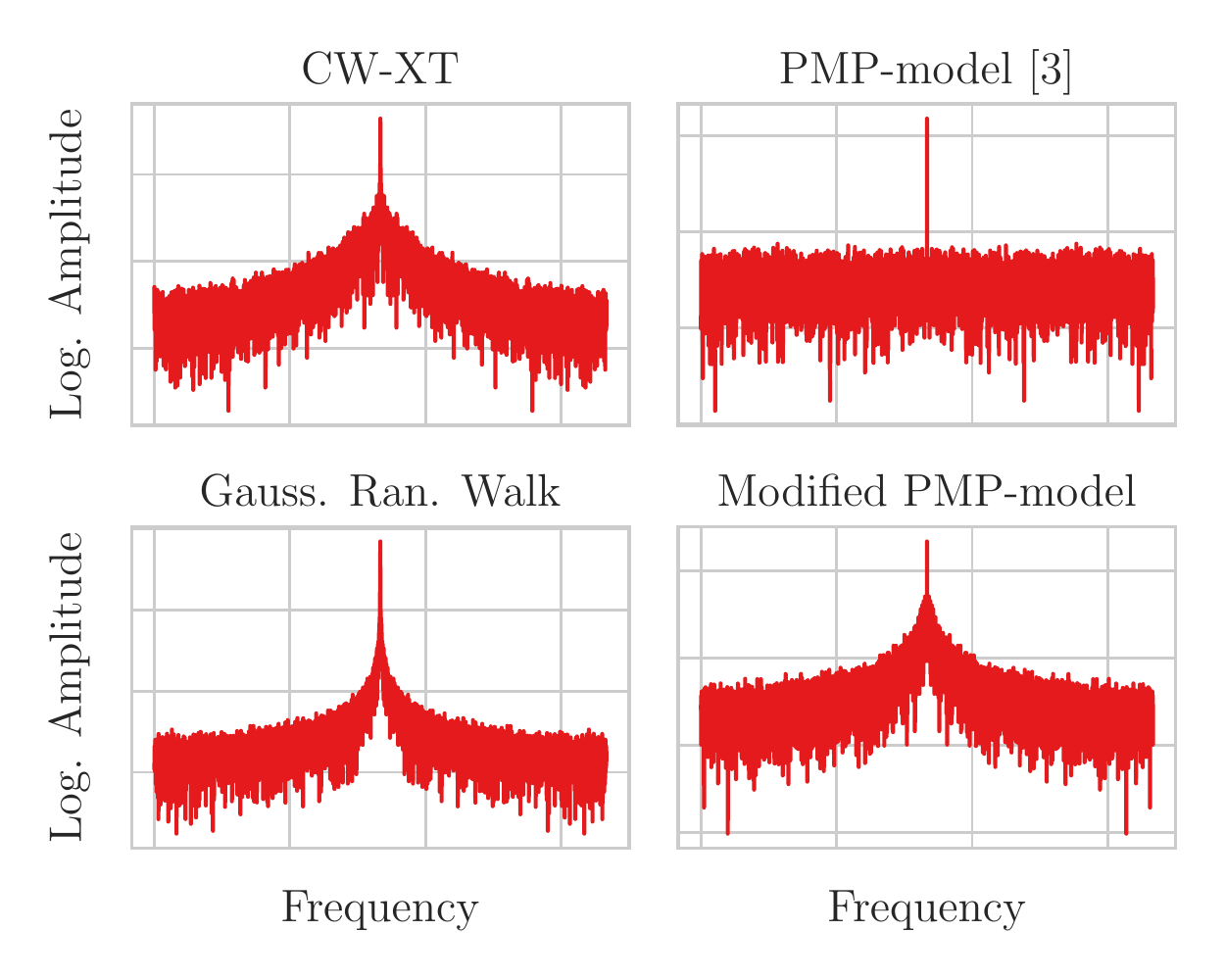}
    \caption{Comparison of FFT of IC-XT and \\theoretical models}
    \label{fig:comp}
  \end{minipage}
\end{figure}
To assess whether a IC-XT can be treated as a random walk, the normalized power spectral density (PSD) of the experimental IC-XT data and that of a simulated Gaussian random walk were compared, showing a clear similarity between the two PSDs (Fig.\ref{fig:PSD}). Additionally, to further prove that the cross-talk can be treated as a random walk in time, the frequency domain representation of the cross-talk and its models have been analysed. As seen in Fig.\ref{fig:comp}, the experimentally measured cross-talk of a continuous wave (CW) signal (top left) largely differs from the one of the model proposed in \cite{Hayashi_2014} (top right) and it has a similar behaviour to a Gaussian random walk (bottom left) and to our proposed modified model of \cite{Hayashi_2014} (bottom right) that is described in the following section, where the $\phi_{\text{rnd},l}$ corresponds to a cumulative sum of random variables in time.\\

To summarize, the model proposed in \cite{Hayashi_2014} is capable of representing the time-domain distribution of CW and OOK, however, it requires a very large time-window and is not capable of representing different modulation formats with higher spectral occupation, probably due to the fact that it would need very large measurement time-windows. In addition, this model is not capable of fully representing the frequency domain behaviour of the IC-XT. On the other hand, analytical analysis such as the one described in \cite{Alves_2016, Staxt1} are able to represent the frequency domain behaviour of the IC-XT. However, the inherent mathematical analysis (STAXT \cite{Alves_2018}) is not in concordance with the model proposed in \cite{Hayashi_2014} because it averages subsequent samples, which for the model in \cite{Hayashi_2014} should be independent, causing the IC-XT intensity to follow a normal distribution considering the central limit theorem (CLT).
In this paper, we propose a novel random walk based model for IC-XT which is a generalization of the one of \cite{Hayashi_2014}, to which it will converge as the time-window tends to infinity. The proposed model is almost independent to time-windows (10s of minutes to hours) and is capable of representing the IC-XT of different source signal formats in both time and frequency domain. It is consistent with the mathematical analysis such as STAXT \cite{Alves_2018} because it allows to keep important statistical information when averaging subsequent samples and maintains the same convergence properties. It also carries  increased statistical information compared to previous methods, using smaller time-windows. Because this model is a generalization of the one of \cite{Hayashi_2014}, all the analysis performed on the latter is still valid in this framework. To prove the validity of the model, an analysis in dB of the step distribution with its heuristic modelling was performed using a Pseudo-Voigt Profile (PVP). This showed that this analysis allowed us to carry most of the statistical information using short time windows (fitting accuracy of 88\% for 20 minutes) and its heuristic fitting is valid for all source-signal formats, temperature and signal properties such as baud rate, modulation format and PRBS. To further prove that the proposed model carries more information than traditional statistical methods, a Machine Learning (ML) classifier was developed for core-pair source signal classification purposes. When the ML was trained using PVP, it showed to be more accurate than when trained using other statistical moments. 

\section{Proposed Model}

\begin{figure}[t]
  \begin{minipage}[b]{0.49\textwidth}
  \centering
    \includegraphics[width=\linewidth]{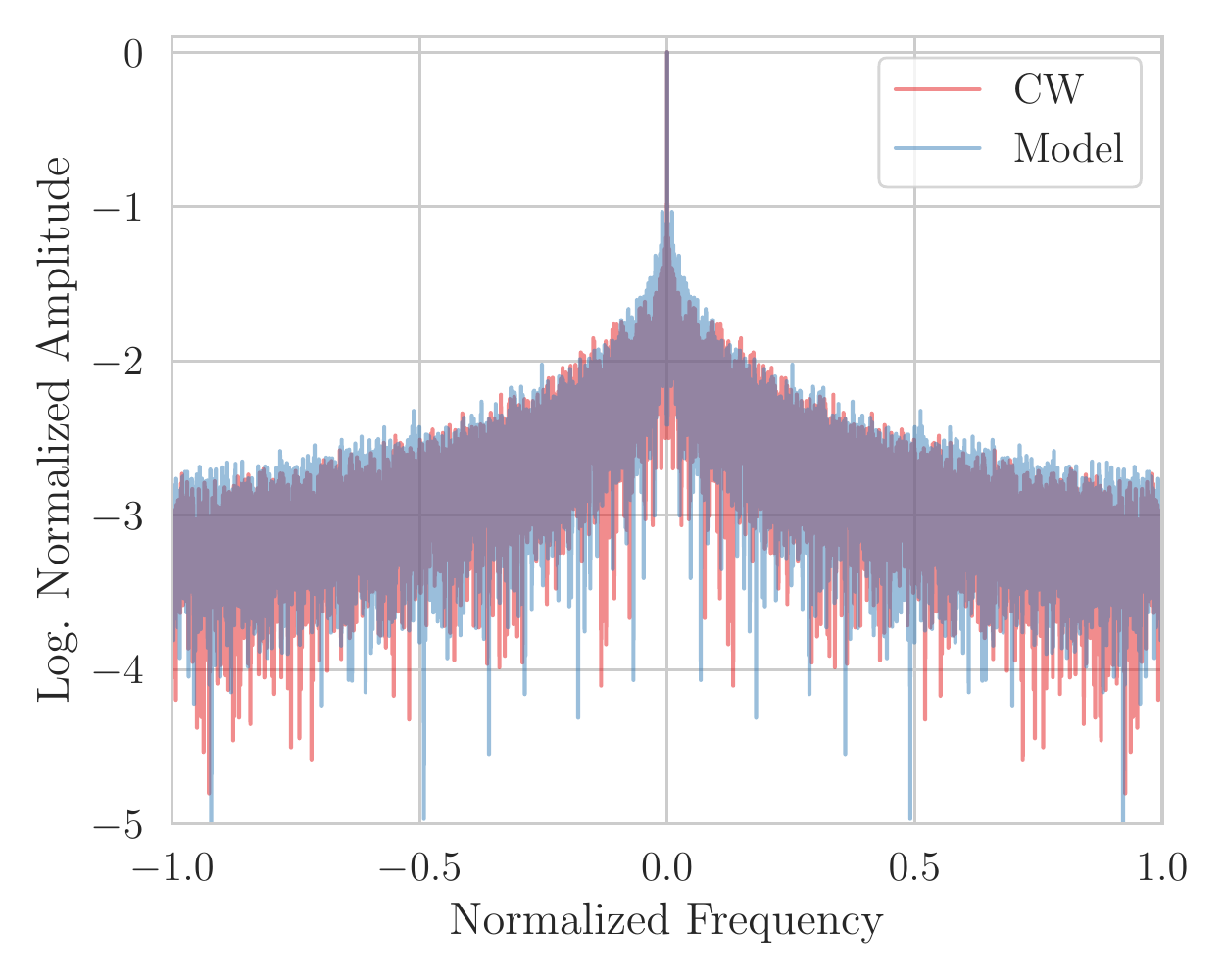}
    \caption{Normalized Log Frequency Response of\\ CW IC-XT and the simulation}
    \label{fig:freq_com}
    \end{minipage}
    \hfill
  \begin{minipage}[b]{0.49\textwidth}
   \centering
    \includegraphics[width=\linewidth]{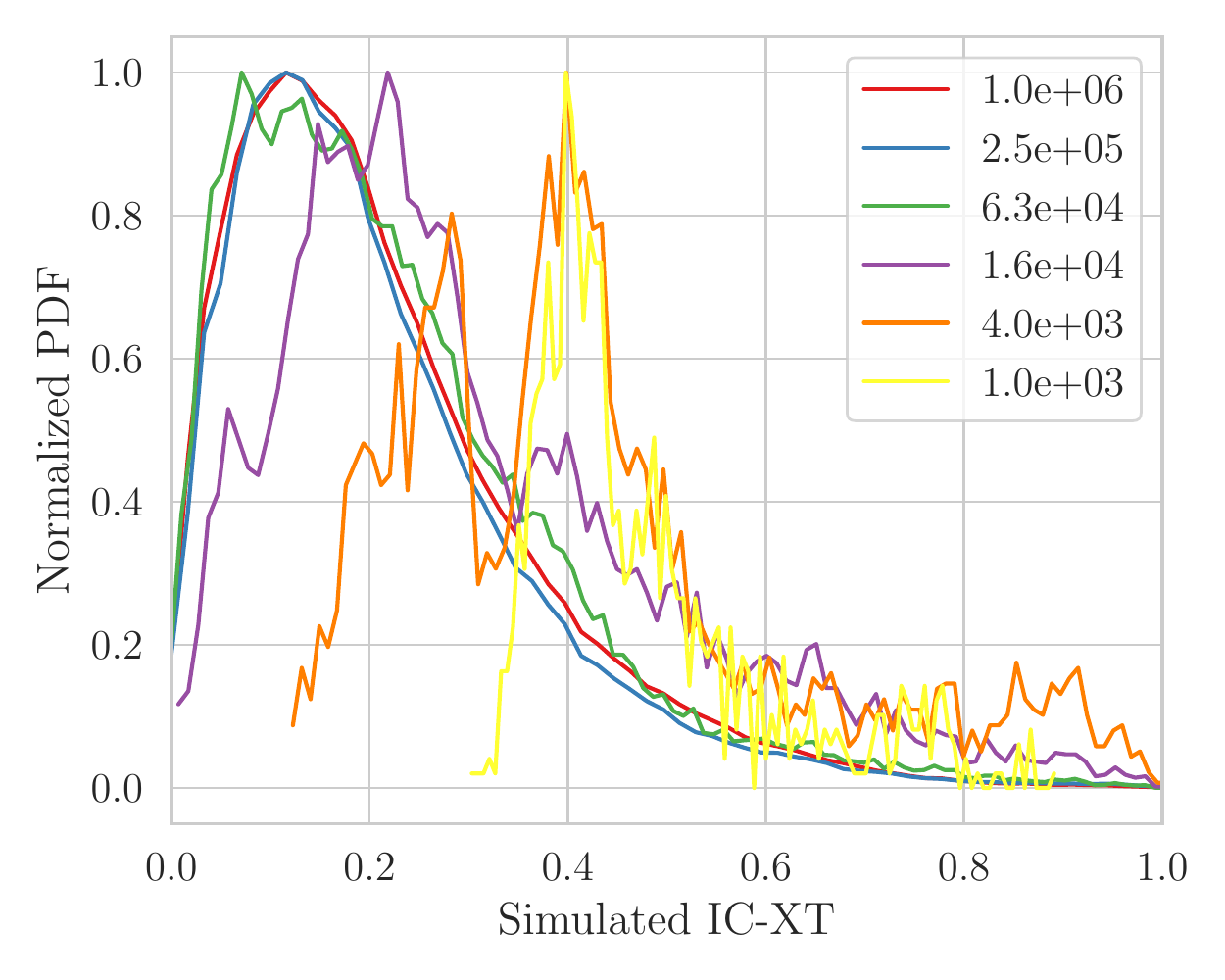}
     \caption{Normalized PDF of IC-XT\\ with different number of samples}
     \label{fig:pdf_sam}
  \end{minipage}
 \end{figure}

We introduce a new model, a generalized version of the \cite{Hayashi_2014}, in which the random phase components are described as random walks in time. 
The model is described as:
\begin{align}
    A_{n,t}(N_{\text{PM}}) &=A_n(0,t) -j \sum_{l=1}^{N_{\text{PM}}} \chi_{nm}(l)\text{exp}\left[-j \phi_{l,t}\right]A_m(l-1)\\
    &\approx -j \sum_{l=1}^{N_{\text{PM}}} \chi_{nm} \text{exp}(-j \phi_{l,t}) \\
    &= -j \sum_{l=1}^{N_{\text{PM}}} \chi_{nm} \text{exp}(-j \phi_{l,t-1} + \gamma)\\
    \label{eq:new_mod}
    &= -j \sum_{l=1}^{N_{\text{PM}}} \chi_{nm} \text{exp}\left[-j \left(\phi_{l,0}+\sum_{k=1}^t\gamma\right)\right]\text{  Where: }\gamma \backsim N(\mu, \sigma^2)
\end{align}
Where $\phi_{l,0}$ are the theoretical phase shifts between the active and target core at the $l^{th}$ phase matching point derived from the equations described in \cite{Hayashi_2010}, $\mu$ and $\sigma$ are the mean and standard deviation of the Gaussian distributed random variable $\gamma$ respectively.
This model is consistent with both time and frequency domain representation of the cross-talk described in Sections \ref{subsec-Time} and \ref{subsec:Freq} . As it can be seen in Figure \ref{fig:freq_com}, the frequency domain representation of the experimental IC-XT clearly resembles the one of the newly proposed model. 

The model proposed can be considered to be a generalization of the one proposed by \cite{Hayashi_2014}, the latter can be considered to be the special case when the time-window and number of samples tends to infinity (proof given in Section \ref{sec:appendix}).  The time needed for the model to converge to the statistical behaviour described by Eq.\ref{chi_dist} strictly depends on the $\mu$ and $\sigma$ parameters of the model and it is out of the scope of this journal. The larger the $\sigma$ the faster it converges. This paper will only consider the case where $\mu = 0$. In addition, this formulation of IC-XT is consistent with other analysis of IC-XT, namely STAXT \cite{Alves_2018} since it allows to keep most statistical information when averaging subsequent samples whilst still converging to the model of \cite{Hayashi_2014}. Another point is with regards to the correlation of IC-XT with source signal spectral occupation, as shown in \cite{Puttnam_2019}, with the  model proposed, the spectral occupation is directly correlated to the $\sigma$ value: the broader is the spectral occupation, the smaller is the $\sigma$ value.

 As it can be seen in Figure \ref{fig:pdf_sam}, before converging, the time domain distribution of the CW signal will follow a multi-modal behaviour, similar to the one seen in Figure \ref{fig:chi_fit}. For this reason, it is not always possible to  analyse the full time domain distribution, however, due to the inherent random walk described by the random variable $\gamma$, we can analyse the distribution of the difference between subsequent samples (step distribution). This analysis was performed to prove that the proposed model is valid. We can prove that most of the information is carried by the characteristics of the stochastic random variable, $\gamma$ and its transformation properties, which can only be found using the difference between subsequent samples. Due to the fact that the step distribution of IC-XT intensity in respect of the parameters of $\gamma$ does not seem to have an analytical solution, an alternative analysis was performed in the dB domain using a heuristic approach, simplifying the problem to a pseudo-random walk.


\section{Step Analysis}
For the aforementioned reasons, we decided to analyze the IC-XT as a pseudo-random walk (Markov process), using the following formulation:
\begin{equation}
\label{eq:Pseudo_random_walk}
S_{t+1} = S_0 + \sum_{n=1}^{n=t} \zeta\quad \text{and} \quad S_{t+1}  = S_{t} + \zeta \quad \quad \text{where}  \quad \zeta \backsim p(\zeta \lvert S_{t})
\end{equation}

\noindent in which $S_t$ is the IC-XT value at a certain time. This process reduces the analysis to a single stochastic random variable $\zeta$ (the difference between every adjacent pair of IC-XT samples at a discrete time interval termed the 'IC-XT step').
The IC-XT is not treated as a pure random walk due to its inherent bounded nature, which depends on the physical fiber structure following mode coupled theory \cite{Hayashi_2011a}. To simplify the analysis, we model the PDF of $\zeta$ over the overall support, and it is found that it can be expressed as:

\begin{equation}
\label{eq:PVP}
p(\zeta)  = \int_{-\infty}^{\infty} p(\zeta \lvert S_{t})p(S_{t})\, d{S_{t} \approx PVP(\zeta;\mu, \sigma, \alpha)},
\end{equation}

 \noindent where PVP is the Pseudo-Voigt profile (a numerical approximation of a Voigt profile). Voigt Profile is a convolution between a Cauchy-Lorentz distribution and a Gaussian distribution~\cite{Pagnini_2018}. PVP consists of a weighted sum between a Gaussian and a Cauchy-Lorentz distribution with the same mean $\mu$ and different standard deviations:
 
\begin{equation}
PVP(\zeta;\mu, \sigma, \alpha) = \frac{\left(1-\alpha\right)}{\sigma_g \sqrt{2 \pi}}e^{\left[-\left(x-\mu\right)^2/{2{\sigma_g}^2}\right]} + \frac{\alpha}{\pi}\left[\frac{\sigma}{\left(x-\mu\right)^2+\sigma^2}\right].
\end{equation}

\begin{figure}[t]
  \begin{minipage}[b]{0.49\textwidth}
  \centering
    \includegraphics[width=\linewidth]{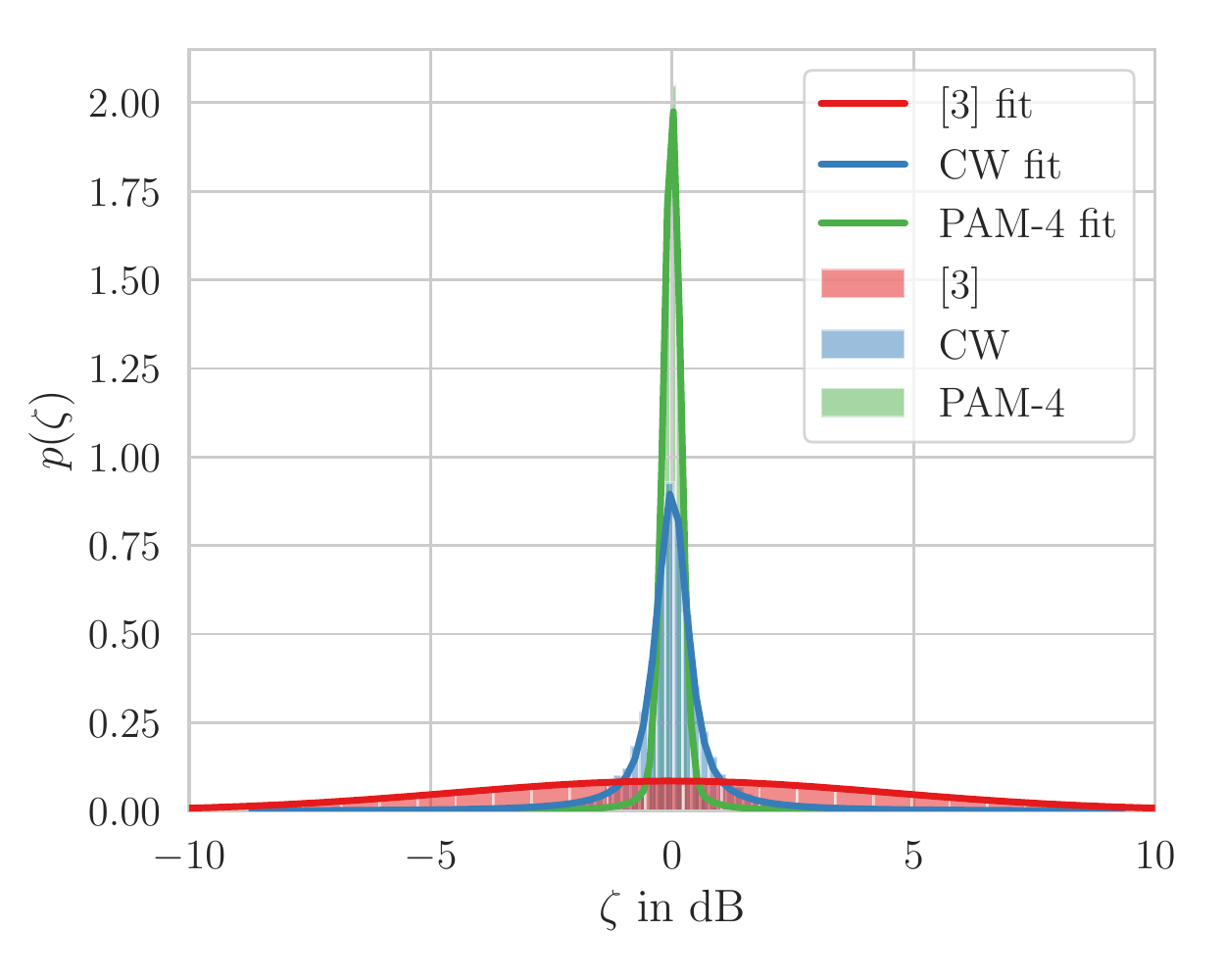}
    \caption{Comparison between step distributions\\
    and PVP fitting}
    \label{fig:step_comp}
    \end{minipage}
    \hfill
  \begin{minipage}[b]{0.49\textwidth}
   \centering
    \small
   \begin{tabular}{c|c|c|c|c}
\rule{0pt}{19pt} \textbf{Source} & $\mu$ & $\sigma$ & $\alpha$ & \footnotesize{Fitt. Acc.} \\ \hline
\rule{0pt}{18pt} \textbf{CW} & -0.088 & 0.371 & 0.840 & 99.56\% \\
\rule{0pt}{18pt} \textbf{ASE} & -0.005 & 0.044 & 0.200 & 99.33\% \\
\rule{0pt}{18pt} \textbf{OOK} & -0.021 & 0.043 & 0.833 & 99.85\% \\
\rule{0pt}{18pt} \textbf{PAM-4} & -0.062 & 0.214 & 0.284 & 99.74\% \\
\rule{0pt}{18pt} \textbf{256-QAM} & 0.367 & 0.004 & -0.0 & 99.87\%
\end{tabular}
\vspace{0.1cm}
      \captionof{table}{Step distribution fitting accuracy and\\coefficients for 12-hour time window}
      \label{tab:fitting_acc_12hr}
  \end{minipage}
 \end{figure}
\noindent The first part of the equation relates to the Gaussian distribution with standard deviation $\sigma_g = \sigma/\sqrt{2 \text{ln} 2}$ and the second to the Cauchy-Lorentz distribution with standard deviation $\sigma$. $\alpha$ is the scaling coefficient between the two distributions (Gaussian $0\leq \alpha \leq 1$ Cauchy). This method has been chosen because both Gaussian and Lauretzian behaviours appeared in the analysis of the experimental data and the proposed method achieves a great fitting accuracy with a relatively small number of parameters.
This choice has also been backed by the fact that a PVP greatly resembles the theoretical step distribution of the IC-XT. In fact, PVP seems to be a almost perfect fit ($R^2$-score$ > 99.95\%$) for this function which does not seem to have a closed form description in the dB domain. One of the main reasons for which the analysis was done in dB, apart for the fact that the problem is not anymore bounded between 0 and $\infty$, is that, assuming that the model of \cite{Hayashi_2014} perfectly represents the IC-XT, the step distribution in the dB domain will always be the same, independently of the $\sigma$ parameter of \ref{chi_dist}. Therefore, the step distribution should not differ for different signal formats. As seen in Figure \ref{fig:step_comp}, this is clearly not the case and IC-XT step distribution in dB arising from different source signal formats differ from each other and from the one expected from \cite{Hayashi_2014}. This is due to the fact that it considers the subsequent samples independent, further proving that the model in \cite{Hayashi_2014} is not capable of thoroughly representing the time-domain behaviour of the IC-XT. Since the PVP has a higher degree of freedom, it is better suited to represent these dependencies which vary depending on various source signal parameters (PRBS length, temperature, format, etc.).

\section{Discussion}

\begin{figure}[t]
  \begin{minipage}[b]{0.49\textwidth}
  \hspace{-0.3cm}
 \centering
    \includegraphics[width=\linewidth]{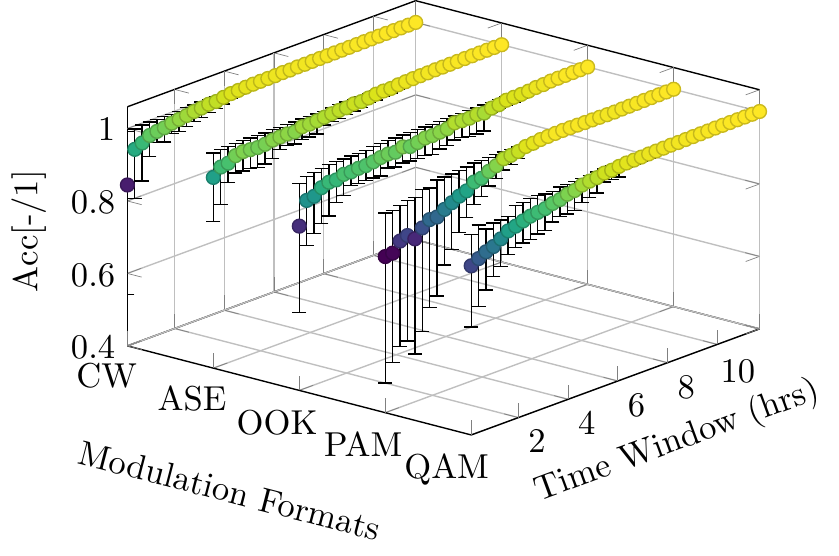}
    \caption{Average and CI$_{90\%}$ Fitting Accuracy\\ vs Time window}
    \label{fig:3D}
  \end{minipage}
  \hfill
  \begin{minipage}[b]{0.49\textwidth}
  \renewcommand\thetable{4}
    \centering
    \small
\hspace{-0.55cm}
\begin{tabular}{l|rrrr}
\rule{0pt}{12pt} \textbf{} & \footnotesize{\textbf{QPSK}} & \footnotesize{\textbf{16-QAM}} & \footnotesize{\textbf{64-QAM}} & \footnotesize{\textbf{256-QAM} }\\ \hline
\rule{0pt}{14pt} \textbf{15 G} & 99.7\% & 99.6\% & 99.2\% & 99.8\% \\
\rule{0pt}{12pt} \textbf{30 G} & 99.5\% & 99.3\% & 99.7\% & 99.7\% \\
\rule{0pt}{12pt} \textbf{45 G} & 99.8\% & 99.6\% & 99.1\% & 99.8\% \\
\rule{0pt}{12pt} \textbf{60 G} & 99.8\% & 100.0\% & 89.9\% & 98.9\% \\
\rule{0pt}{12pt} \textbf{80 G} & 95.9\% & 97.7\% & 95.4\% & 97.2\%
\end{tabular}
\vspace{0.05cm}
\captionof{table}{$R^2$ score for different Baud rates\\and modulation formats}
      \label{tab:baud_mod}
  \end{minipage}
\end{figure}

The proposed model has been fitted to the IC-XT step distribution of the experimental data (from \cite{Yuan_2019}) with different source signal formats within a 12-hour time window. As shown in Fig.\ref{fig:step_comp} and Table \ref{tab:fitting_acc_12hr}, the step distributions for different signalling sources closely fit the proposed model (> 99.3\% fitting accuracy). To verify the validity of the proposed model under different signalling conditions, the steps of one hour of measured QAM data with different modulation formats (from QPSK to 256-QAM) and symbol rates (from 20GBaud/s to 80GBaud/s) have been fitted Table \ref{tab:baud_mod}. These lead to an average accuracy of 98.4\% with CI$_{90\%}$ between 95.1\% and 99.8\%. A similar analysis has been performed for one hour OOK data with different PRBS patterns and temperatures, Table \ref{tab:PRBS temperature acc}, leading to an average fitting accuracy of 99.4\% with CI$_{90\%}$ between 99.0\% and 99.8\%. Temperature analysis on ASE under different temperatures ($20^\circ$C to $50^\circ$C) lead to average fitting accuracy 98.4\% with CI$_{90\%}$ between 98.0\% and 98.9\%. The results show that the proposed model is capable of representing all possible signalling and measurement conditions. 

\begin{table}[b]
\renewcommand\thetable{3}
    \centering
    \small
\begin{tabular}{l|rrrrrrrr}
 Temperature & \multicolumn{8}{c}{PRBS pattern} \\
\textbf{} & \textbf{7} & \textbf{9} & \textbf{10} & \textbf{11} & \textbf{15} & \textbf{20} & \textbf{23} & \textbf{31} \\ \hline
\textbf{30\textdegree C} & 99.1\% & 99.5\% & 99.3\% & 99.3\% & 99.5\% & 99.1\% & 99.4\% & 99.3\% \\
\textbf{40\textdegree C} & 99.1\% & 99.6\% & 99.5\% & 99.2\% & 99.7\% & 99.5\% & 99.8\% & 99.6\% \\
\textbf{50\textdegree C} & 99.6\% & 99.9\% & 99.8\% & 99.4\% & 99.6\% & 98.9\% & 99.2\% & 99.2\%
\end{tabular}
\label{tab:PRBS temperature acc}
      \captionof{table}{$R^2$ Score fitting accuracy for different temperature and PRBS}
      \label{tab:PRBS temperature acc}
    \end{table}

Furthermore, the resilience of the model is evaluated. The model was fit to increasingly large subsets of the 12 hours of data, starting at 6 minutes and increasing in intervals of 18 minutes up to the full set of 12 hours. The fitting procedure was applied multiple times at each interval on different subsets of data, and the fitting accuracy evaluated. This type of analysis shows how much of the full information can be retrieved with smaller time windows. Fig.\ref{fig:3D} shows the average accuracy and the CI$_{90\%}$ for every source signal with respect to the time window. Note that the fitting accuracy in the figures refers to the similarity of the fitted PVP distribution of current samples to that of the benchmark samples of each source, which is also evaluated through the $R^2$ score function. A benchmark sample corresponds to the fitted model for a specific signal source data with a time window of twelve hours.

\begin{table}[b]
\centering
\small
\begin{tabular}{l|l|l|ll|l|l}
 & \multicolumn{2}{c|}{Resampled} & \multicolumn{2}{c|}{Original} & \multicolumn{2}{c}{Res. + Ori.} \\ \cline{2-7} 
\multicolumn{1}{c|}{} & \multicolumn{1}{c|}{Validation} & \multicolumn{1}{c|}{Test} & \multicolumn{1}{c|}{Validation} & \multicolumn{1}{c|}{Test} & \multicolumn{1}{c|}{Validation} & \multicolumn{1}{c}{Test} \\ \hline
Moments & 58.6$\pm$4.7\% & 45.16\% & \multicolumn{1}{l|}{61.7$\pm$5.0\%} & 48.39\% & 59.4$\pm$13\% & 58.06\% \\
Fitted Coeff. & 68.7$\pm$4.1\% & 77.40\% & \multicolumn{1}{l|}{76.0$\pm$6.0\%} & 90.32\% & 80.2$\pm$8.1\% & 87.10\% \\
Mom.+Coeff. & 62.2$\pm$18\% & 74.50\% & \multicolumn{1}{l|}{78.2$\pm$3.5\%} & 82.35\% & 75.5$\pm$13\% & 82.16\%
\end{tabular}
\caption{SVM source signal classification accuracy}
\label{tab:classify}
\end{table} 
Even with six minutes of data, the model achieves an average accuracy above 83\% for every modulation format, and for a time window above 4 hours the CI$_{90\%}$ is above 90\%. These results suggest that the proposed technique is able to convey most relevant information using a short observation time, making it a suitable candidate for IC-XT characterization, source signal identification and classification. An empirical analysis regarding this matter has been performed using one hour long samples of IC-XTs arising from different source signals and measured with different averaging times (0.025 to 3.5 seconds). For this analysis, the main statistical moments (mean, variance, skewness and kurtosis) of these samples and the PVP coefficients have been calculated. These elements have then been normalized and used as inputs of a support vector machine (SVM) classifier which had as labels the corresponding source signaling formats. Different classifiers have been trained in different conditions: a) using only statistical moments, b) using only fitting coefficients of the proposed PVP model, c) using a combination of both. Also we use original data, re-sampled data and combination of them to simulate the same averaging time, Table.\ref{tab:classify}. For each possibility, a different classifier has been trained and found (the most accurate under cross-validation) using grid search cross-validation\cite{GridSearchCV} over 64 possible hyper-parameter combinations, using a data train-test split of 70-30\%, and validated using a six-fold Stratified K-fold cross-validator \cite{Stratified_k_fold}. The results of this analysis are outlined in Table \ref{tab:classify}.

As seen from Table \ref{tab:classify}, the classification accuracy is best when only the PVP fitting coefficients are considered. This proves that these elements carry most of the useful information and that the other statistical methods are too noisy to lead to any favorable conclusion. In this analysis,  the fitting coefficients of the chi-square model of \cite{Hayashi_2014} were discarded due to the fact that the $R^2$ score of the fitting were below 10\%. 
It is important to note that the developed classifier cannot be considered a general tool for source signal classification due to the relatively small amount of data and large variation in measurement conditions on which it has been trained and validated. Nonetheless, our shown classifier is a valid proof of concept that the pseudo-random walk representation of the cross-talk is capable of carrying more information than other proposed solutions.

\section{Conclusion}

In this paper, a novel random walk based model for IC-XT which is capable of representing the IC-XT of different source signal formats in both time and frequency domain was proposed. It has been proven that this model is a generalized version of the most widely used model in literature, to which it converges. Compared to the previous model \cite{Hayashi_2014} which represents the Ic-XT only when the time window tends to infinity, the proposed model can accurately represent the IC-XT statistical properties independently from the time window, and it creates a framework for in which analytical analysis such as STAXT are valid and consistent with theory because they are not limited by the CLT. In addition, the proposed model, being a generalization of model \cite{Hayashi_2014}, keeps all the previously performed analysis valid. This model was validated using a new pseudo-random walk based characterization in dB domain, capable of fitting various source signal formats ($R^2 > 99.3\%$). This characterisation showed great performance under different measurement conditions (temperature, PRBS, modulation order, baud rate) and great resilience to the time window. Statistical analysis and evaluation of this method have been carried out and its effectiveness in carrying relevant statistical information has been empirically proved using ML techniques for source signal classification. This work highlights the importance of gaining a better understanding of the time and frequency domains of the dynamic cross-talk to be able to perform an accurate characterization of its behaviour. This topic remains relatively unexplored within the optics community and understanding it better could potentially lead to significant improvements in future technologies such as SDM systems using MCF.      

\section{Disclosure}
This work is partly supported by EPSRC TRANSNET grant EP/R035342/1 and Microsoft Research studentship to Alessandro Ottino.

\section{Data and Supplementary Material}

The experimental data from which the conclusion in this paper can be accessed in \cite{Dataset:measurements}. Supplementary material, analysis and SVM hyper-parameters information are available at \cite{ottino_2020}.

\section{Appendix}
\label{sec:appendix}
In Eq. \ref{eq:new_mod} we define the complex amplitude of IC-XT at  $N_{PM}^{th}$ PMP at time t as:
\begin{align}
    A_{n,t}(N_{\text{PM}}) &\approx -j \sum_{l=1}^{N_{\text{PM}}} \chi_{nm} \text{exp}\left[-j \left(\phi_{l,0}+\sum_{k=1}^t\gamma\right)\right]\text{  Where: }\gamma \backsim N(\mu, \sigma^2).
\end{align}
The phase component $\phi_{l,0}+\sum_{k=1}^t\gamma$ is an independent variable of the function $f(x)=exp[-jx]$. The latter function is periodic with period of $2\pi$. This transformation modify the support of phase variable from the whole $\mathbb{R}$ to a circular plane in $\mathbb{R}$ with support $[0,2\pi)$, so a finite group in $\mathbb{R}$. For this reason the problem can be seen as a random walk in $\mathbb{R}_n$, the real modulo $n=2\pi$. This random walk corresponds to a Markov chain which is irreducible, aperiodic and double stochastic (\cite{sinclair, ross_2000}), thus its stationary thus its stationary probability will be uniformly distributed on $\mathbb{R}_n$, and therefore on the support $[0, 2\pi)$. Additional information and proof can be found in: \cite{Diaconis93comparisontechniques, ALDOUS198769,Hildebrand2004ASO}. This translates to:
\begin{align}
     p\left(\lim_{t \to +\infty}\text{exp}\left[-j \left(\phi_{l,0}+\sum_{k=1}^t\gamma\right)\right]\right) = p(\text{exp}[-j \phi_{\text{rnd},l}]) \text{  Where: }\phi_{\text{rnd},l} \backsim U(0, 2\pi)
\end{align}
so:
\begin{align}
     \lim_{t \to +\infty}[A_{n,t}(N_{\text{PM}})] &\approx \lim_{t \to +\infty}\left[-j \sum_{l=1}^{N_{\text{PM}}} \chi_{nm} \text{exp}\left[-j \left(\phi_{l,0}+\sum_{k=1}^t\gamma\right)\right]\right]\\
     &= -j \sum_{l=1}^{N_{\text{PM}}} \chi_{nm} \text{exp}(-j \phi_{\text{rnd},l})\text{  Where: }\phi_{\text{rnd},l} \backsim U(0, 2\pi)
\end{align}
Therefore when the times tends to infinity the model tends to the one proposed in \cite{Hayashi_2014}.

\bibliography{sample}
\end{document}